\documentclass[aps, prl, 10pt, superscriptaddress, twocolumn]{revtex4-1}

\usepackage{amsmath,amssymb,mathrsfs}
\usepackage{graphicx}

\newcommand{\la}{\label}
\newcommand{\bbm}{\begin{multline}}
\newcommand{\eem}{\end{multline}}
\newcommand{\be}{\begin{equation}}
\newcommand{\ee}{\end{equation}}
\newcommand{\bea}{\begin{eqnarray}}
\newcommand{\eea}{\end{eqnarray}}
\newcommand{\p}{\partial}

\newcommand{\cP} {\mathcal{P}}
\newcommand{\cQ} {\mathcal{Q}}
\newcommand{\cK} {\mathcal{K}}
\newcommand{\cZ} {\mathcal{Z}}
\newcommand{\cS} {\mathcal{S}}
\newcommand{\cO} {\mathcal{O}}
\newcommand{\vars} {{ \rm var}(s)}
\newcommand{\dxi} {[d\xi]}
\newcommand{\dxip} {[d\xi^\prime]}
\newcommand{\dz} {[dz]}
\newcommand{\cz} {\{z\}}
\newcommand{\cxi} {\{\xi\}}
\newcommand{\cxip} {\{\xi^\prime\}}
\newcommand{\comment}[1]{}
\usepackage{color}
\def\red{\color{red}}

\allowdisplaybreaks[1]

\begin{document}

\title{\begin{flushright}\vspace{-1in}
			\mbox{\normalsize  EFI-16-28}
		\end{flushright}
	Particle-Hole Duality in the Lowest Landau Level \vskip 20pt
	 }

\author{Dung Xuan Nguyen}
\affiliation{Department of Physics, University of Chicago, Chicago, Illinois 60637, USA}

\author{Tankut Can}
\affiliation{Simons Center for Geometry and Physics, Stony Brook University, Stony Brook, NY
11794, USA}

\author{Andrey Gromov}
\affiliation{Kadanoff Center for Theoretical Physics 
and Enrico Fermi Institute, University of Chicago, Chicago, Illinois 60637}

\date{\today}

%%%%%%%%%%%%%%%%%%%%%%%%%%%%%%%%%%%%%%%%%%%%%%%%%%%%%%%%%%%%%%%%%%%%%%%%%%%%%%%%%%%%%%%%%%%%%%%
\begin{abstract}
We derive a number of exact relations between response functions of holomorphic, chiral fractional quantum Hall states and their particle-hole (PH) conjugates. These exact relations allow one to calculate the Hall conductivity, Hall viscosity, various Berry phases, and the static structure factor of PH-conjugate states from the corresponding properties of the original states. These relations establish a precise duality between chiral quantum Hall states and their PH-conjugates. The key ingredient in the proof of the relations is a generalization of Girvin's construction of PH-conjugate states to inhomogeneous magnetic field and curvature. Finally, we make several non-trivial checks of the relations, including for the Jain states and their PH-conjugates.
\end{abstract}

%\pacs{03.65.Ud}{Entanglement and quantum nonlocality} -- CORRECT
%\pacs{05.30.Fk}{Fermion systems and electron gas}
%\pacs{05.40.-a}{Fluctuation phenomena, random processes, noise, and Brownian motion} -- CORRECT

%%%%%%%%%%%%%%%%%%%%%%%%%%%%%%%%%%%%%%%%%%%%%%%%%%%%%%%%%%%%%%%%%%%%%%%%%%%%%%%%%%%%%%%%%%%%%%%

\maketitle

%%%%%%%%%%%%%%%%%%%%%%%%%%%%%%%%%%%%%%%%%%%%%%%%%%%%%%%%%%%%%%%%%%%%%%%%%%%%%%%%%%%%%%%%%%%%%%%

%%%%%%%%%%%%%%%%%%%%
\paragraph{Introduction.} 
%%%%%%%%%%%%%%%%%%%%

Particle-hole  (PH) transformation for fractional quantum Hall (FQH) states was introduced by Girvin \cite{girvin1984particle}. This transformation relates  a FQH state at filling fraction $\nu$ to a FQH state at filling fraction $1-\nu$. In the absence of Landau level mixing the projected lowest Landau level (LLL) Hamiltonian is PH-symmetric and, therefore, two states related by a PH transformation have the same energy (up to a shift in the chemical potential). Despite the physical clarity of PH-symmetry, the PH-transformed wave functions look quite complicated and are difficult to work with. PH-transformed states contain a different number of particles, have different transport properties and different topological order. In this Letter we will explain that all of the information about PH-transformed state is encoded in the original state, so that both states are a different representation for essentially the same physics. For this reason we feel it is more appropriate to refer to the PH-transformation as a {\it particle-hole duality} (PHD).

Recent years have also brought the rise of interest in the role of PHD in the problem of the half-filled Landau level. To resolve the issue of the apparent absence of the PH-invariance in the Halperin-Lee-Read \cite{halperin1993theory} theory, Son has proposed a manifestly PH-invariant effective theory of composite fermions with $\pi$ Berry phase around the composite Fermi surface \cite{son2015composite}. This theory can successfully be used to describe Jain states at fillings close to $\nu=1/2$ and a PH-invariant (or self-dual) version of the Pfaffian state  \cite{chen2014symmetry, son2015composite}, which is a viable candidate for the observed $\nu = 5/2$ plateau \cite{zucker2016particle}.

PH-transformation, as defined by Girvin \cite{girvin1984particle}, works in flat space and homogeneous magnetic field. It was recently appreciated that placing a FQH state in inhomogeneous background magnetic field and curved geometry allows one to extract  considerable information about the flat space properties of the state \cite{WenZeeShiftPaper, lee1994orbital,haldane2009hall, klevtsov2014random, Abanov-2014, Gromov-galilean, gromov-thermal, gromov2016boundary, GCYFA, bradlyn2014low,  bradlyn2015topological,  Klevtsov-fields, klevtsov2015precise, CLW,can2014geometry,laskin2015collective,laskin2016emergent,klevtsov2015quantum, 2014-ChoYouFradkin, hughes2011torsional}. For example, the projected static structure factor (SSF) \cite{1986-GirvinMacDonaldPlatzman} in leading and sub-leading order in momentum, and long-wave corrections to Hall conductivity and Hall viscosity can be calculated from the properties that become apparent in curved space \cite{CLW, bradlyn-read-2012kubo, 2012-HoyosSon, Gromov-galilean, son2013newton}. Integer quantum Hall states in curved geometry are available in (synthetic) photonic systems \cite{schine2015synthetic}.

In this Letter, we will use the approach of \cite{klevtsov2014random, CLW} to extend Girvin's construction to inhomogeneous magnetic field and curved geometry. Next, we will derive several exact relations between Hall conductivity, Hall viscosity, Berry phases, and the SSF of the holomorphic, chiral FQH states and their PH-duals. These relations establish the PHD quantitatively and show that properties of the PH-dual state are completely determined by the original state. The duality is non-trivial since the calculations can be easily done before the PH-transformation, but are difficult to do after.

Under certain assumptions, the long-wave corrections to Hall conductivity, Hall viscosity and the SSF are determined by topological quantum numbers  \cite{2012-HoyosSon, bradlyn-read-2012kubo, Gromov-galilean, CLW, can2014geometry}: filling fraction $\nu$, shift \cite{WenZeeShiftPaper} $\mathcal S=2\bar s$, chiral central charge \cite{kitaev2006anyons} $c_-$, and the orbital spin variance \cite{GCYFA, bradlyn2015topological} ${ \rm var}(s)$. We will explain how the topological quantum numbers transform under the PHD and prove that the aforementioned long-wave corrections are still determined by the (transformed) topological quantum numbers, albeit via different relations. We will check the derived relations against the explicit computation of the corresponding quantities for Jain states done in Son's theory of composite fermions and find complete agreement.

%%%%%%%%%%%%%%%%%%%%%%%%%%%%
\paragraph{FQH states in inhomogeneous background.} 
%%%%%%%%%%%%%%%%%%%%%%%%%%%%

We start with a brief review of the construction \cite{klevtsov2014random, CLW,can2014geometry} of a FQH state in inhomogeneous magnetic field and curvature. Consider a holomorphic FQH state $\Psi_\nu(\cxi)$, where $\cxi=\xi_1,\ldots,\xi_N$ denotes the collection of particle coordinates. Complex variable $\xi= x+ iy$ will be used to label the particle position in the plane. We will assume that the magnetic field $B$ is inhomogeneous and the background geometry is curved. Then the  unnormalized wavefunction $\Psi_\nu(\cxi)$ takes the following form \cite{klevtsov2014random, CLW}
\be\la{ansatzWF}
\Psi_\nu(\cxi) =  f_\nu(\cxi) e^{\frac{1}{2} \sum_{i=1}^N \cQ (\xi_i,\bar \xi_i)}\,,
\ee
where $\cQ$ is the magnetic potential \footnote{We have chosen to use the natural units \protect{$e=c=\hbar=1$}} defined by
\be\la{MP}
\Delta_g \cQ = - 2B\,,
\ee
where $\Delta_g$ is the Laplace operator for the metric $g_{ij}$. Throughout the Letter we will fix the coordinates so that $g_{ij} = \sqrt{g} \delta_{ij}$. In these coordinates (also known as the ``conformal gauge'') the Laplacian is given by $\Delta_g = \frac{4}{\sqrt{g}}  \p_{z} \p_{\bar{z}}$,. When the magnetic field is homogeneous, but the space is curved the magnetic potential is given by
\be\la{KP}
\cQ = - \frac{\cK}{2\ell^2}\,,
\ee
where $\cK$ is the  K\"ahler potential satisfying $ \p_{z}  \p_{\bar{z}} \cK = \sqrt{g}$, and $\ell = B^{-1/2}$ is the magnetic length. It is of crucial importance that $f_\nu(\cxi)$ {\it does not depend on}  $\cQ$ or the metric $\sqrt{g}$. This will not be the case for PH-dual states.

The central object of interest is the logarithm of the normalization factor 
\be\la{genfun_nu}
 \cZ_\nu[W] =  \int \dxi |f_\nu(\cxi)|^2 \, e^{\sum_{i}W(\xi_i, \bar{\xi}_i)}\,,
\ee
where  $\dxi = d^{2}\xi_1\cdots d^{2}\xi_N$ with $d^{2} \xi = dx dy$, and $W = \cQ + \log \sqrt{g}$. We assume that for constant magnetic field and flat space, when $W = -|z|^{2}/2\ell^{2}$, the state is normalized and $\mathcal{Z_\nu} = 1$. 
 It is not hard to see that $ \log \cZ_\nu[W]$ is the generating functional of the density correlation functions \cite{CLW} 
\be\la{GenF}
\langle \rho_{\nu}(\zeta) \rangle  \equiv \langle \Psi_{\nu}| \rho_\nu(\zeta)| \Psi_{\nu} \rangle = \frac{1}{\sqrt{g}}\frac{\delta \, \log \cZ_\nu[W]}{\delta W(\zeta)}\,,
\ee
where  $\rho_\nu(\zeta)= \frac{1}{\sqrt{g}} \sum_{i=1}^{N_{\nu}} \delta(\zeta - \xi_i)$  is the density operator, and  $N_{\nu}$ is the number of particles in the state $\Psi_{\nu}$. In writing $\langle \rho_\nu \rangle$ we will always implicitly assume that the expectation value is taken in the state with the filling factor $\nu$.

The second variation produces the connected two-point function \cite{CLW}
\be\la{2ptden}
\langle \rho_\nu(\zeta)\rho_\nu(\zeta^\prime) \rangle_{c}  =\frac{\delta }{\delta W(\zeta^\prime)}\langle\rho_\nu(\zeta)\rangle\,,
\ee
where $\langle \rho_\nu(\zeta) \rho_\nu(\zeta') \rangle_{c} = \langle \rho_\nu(\zeta) \rho_\nu(\zeta') \rangle - \langle \rho_\nu(\zeta) \rangle \langle \rho_\nu(\zeta') \rangle$. The static structure factor (SSF) is defined as the Fourier transform
\be \la{SSF} 
 s_{\nu}(q) =   \frac{1}{\bar{\rho}_{\nu}}\langle \rho_\nu(q) \rho_\nu(-q)\rangle_{c}\,,
\ee
where  $\bar \rho_\nu = \nu/(2\pi \ell^{2})$  is the mean electron density in the homogeneous limit (in the bulk of the FQH droplet).

It follows from \eqref{MP} that in flat space, derivatives w.r.t. $W$ and $B$ can be traded with each other. Going to momentum space, using the dimensionless momentum $|k\ell| = q$ we recover \cite{CLW}

\be\la{SSFsigma}
 s_\nu(q) =  \frac{q^2}{2} \frac{2\pi}{\nu}\frac{\delta \langle\rho_\nu(q) \rangle}{\delta B(-q)} =\frac{q^2}{2}\frac{\sigma^H_\nu(q)}{\sigma_{\nu}^{H}(0)}\,,
\ee
where we used the  St\v reda formula $\delta \langle\rho_\nu\rangle/\delta B =  \sigma_\nu^{H}$ \cite{Streda:1982qf},  and the DC Hall conductance $\sigma_{\nu}^{H}(0) = \nu/ 2\pi$.

We will also need to know how the electron density depends on the spatial curvature. This dependence is captured by the function $\eta_\nu(q) = \frac{2\pi}{\nu} \delta \langle \rho_\nu\rangle /\delta R$. In general, $\eta_\nu(q)$ has the following momentum expansion 
\be
\eta_\nu(q) = \frac{\cS}{4} - \frac{b}{4\nu} q^2 + O(q^4)\,,
\ee
where the constant $b$ is an {\it a priori} non-universal parameter. However in the LLL it is determined by the topological quantum numbers \cite{Gromov-galilean} 
\be\la{buniversal}
b = \nu \bar s\left(1-\bar s\right)  +\frac{\tilde c}{12}\,,
\ee
where $\tilde c=c_- -  12\nu \,{\rm var}(s) $. 
It is also known to control the Berry curvature on the moduli space of higher genus surfaces \cite{klevtsov2015precise}.
 Note that the kinematic Hall viscosity \cite{1995-AvronSeilerZograf} follows from the zero momentum limit of the curvature response $\eta^{H}_\nu/\bar \rho_{\nu} = \eta_{\nu}(0)$ \cite{can2014geometry}.
At the same time using the expression for the scalar curvature $R = - \Delta_g \ln \sqrt{g}$, and the general relation, valid for any metric-independent operator $\cO$ \cite{CLW} 
\be\la{g-W}
- \frac{\ell^2}{2} \Delta_g \frac{\delta \langle \mathcal O \rangle}{\delta \sqrt{g}} = \left(1 - \frac{\ell^2}{2} \Delta_g\right)  \langle \mathcal{O} \rho_\nu \rangle_{c}
\ee
it is possible to show that \cite{CLW}
\be\la{SSFeta}
s_\nu(q) = \frac{q^2/2}{1+q^2/2} \left(q^2 \,\eta_\nu(q) + 1\right)\,.
\ee
These relations imply
\be 
s_{\nu}(q) = \frac{1}{2} q^{2} + \frac{\mathcal{S} - 2}{8} q^{4} -  \left[ \frac{b}{8\nu} + \frac{(\mathcal{S} - 2)}{16} \right]q^{6} + \ldots\,.
\ee 
%
%\be
%\bar s_{\nu}(q) = \frac{\cS-1 }{8}q^4 + \frac{6b+5\nu- 3\nu\cS}{48\nu}q^6  +\ldots
%\ee
Finally, combining \eqref{SSFsigma} and \eqref{SSFeta} we establish an exact relation between Hall conductivity and $\eta_\nu(q)$
\be\la{sigma-eta}
 \sigma_{\nu}^{H}(q) = \frac{\nu}{1+q^2/2} \left(q^2 \eta_\nu(q) + 1\right)\,.
\ee
Relations \eqref{SSFsigma}, \eqref{SSFeta} and \eqref{sigma-eta} hold to all orders in $q$, under the assumption of the absence of Landau level mixing and long-range interactions. Together with \eqref{buniversal} these relations imply that first 3 terms in the momentum expansion of  $\sigma^H_\nu(q)$ and $s_\nu(q)$ are completely determined by the topological quantum numbers. 
%%%%%%%%%%%%%%%%%%%%%%%%%%%%
\paragraph{Particle-hole transformation in inhomogeneous background.} 
%%%%%%%%%%%%%%%%%%%%%%%%%%%%

Following Girvin \cite{girvin1984particle}, we define $\Psi_{1-\nu}(\cz)$ as a state of {\it holes} at filling $\nu$, which, when viewed as a state of electrons, has filling $1-\nu$. Let $z_1,\ldots,z_M$ be the coordinates of electrons and $\xi_1,\ldots,\xi_N$ be the coordinates of holes. Then the PH dual state is defined as
\be\la{PH-def}
\Psi_{1-\nu}(\cz) = \sqrt{\frac{(N+M)!}{N!M!}} \int \dxi  \Psi_1(\cz,\cxi)\Psi_\nu^*(\cxi)\,,
\ee
where $\Psi_\nu(\cxi)$ is given by \eqref{ansatzWF} and $\Psi_{1}$ is the $\nu = 1$ state. The overall factor is required to ensure that the PH-dual state is normalized to $1$ in constant magnetic field and flat space. A defining property of the PH transformation is that it is an involution 
\be\la{inv}
\Psi_{1-(1-\nu)}(\cxi) = \pm \Psi_\nu(\cxi)\,.
\ee
\begin{widetext} 
Property \eqref{inv} is ensured by the following identity. First, we define an $n$-particle reduced density matrix \cite{macdonald1988density}
 \be \label{densitymatrix}
 \cP_\eta^{(n)}(\xi_1,\cdots,\xi_n;\xi'_1,\cdots,\xi'_n) = \frac{N_{\eta}!}{n! (N_\eta-n)!} \frac{1}{\mathcal{Z}_{\eta}[W]}\int [d\hat{\xi}] \Psi_\eta(\hat{\xi}_{n+1},\cdots,\hat{\xi}_{N_\eta},\xi_1,\cdots,\xi_n)\Psi^*_\eta(\hat{\xi}_{n+1},\cdots,\hat{\xi}_{N_\eta}, \xi'_1,\cdots,\xi'_n)\,.
\ee
%\end{widetext} 
 This density matrix is a projector to the LLL satisfying
\be\la{projector}
\Psi_\nu(\cxi) = \int \dxip \cP_1^{(N_{\nu})}(\cxi,\cxip)  \Psi_\nu(\cxip)\,.
\ee
Some properties of $\cP_\nu^{(n)}(\cxi,\cxip)$ as well as its explicit form are discussed in the Supplementary Material.  In particular, we prove the following formula relating the 2-particle reduced density matrices between PH-dual states 
\bea\nonumber
\cP_{1-\nu}^{(2)}(\xi_1,\xi_2; \xi_1,\xi_2) &&= 	 \cP_{1}^{(2)}(\xi_1,\xi_2; \xi_1,\xi_2)+\cP_{\nu}^{(2)}(\xi_1,\xi_2; \xi_1,\xi_2) +   \frac{1}{2}\cP_{1}^{(1)}(\xi_1; \xi_2)\cP_{\nu}^{(1)}(\xi_2;\xi_1) + \frac{1}{2}\cP_{1}^{(1)}(\xi_2;\xi_1)\cP_{\nu}^{(1)}(\xi_1;\xi_2)\\ \label{exact_density}
&&  - \frac{1}{2}\langle \rho_{1}(\xi_1)\rangle \langle \rho_{\nu}(\xi_2)\rangle - \frac{1}{2}\langle\rho_{1}(\xi_2)\rangle\langle \rho_{\nu}(\xi_1)\rangle\,.
\eea
\end{widetext} 
Integrating over position $\xi_2$ reduces this to a simple formula relating the electron density (in inhomogeneous background)
\be\la{density}
 \langle \rho_\nu \rangle + \langle\rho_{1-\nu}\rangle = \langle \rho_1\rangle \,.
\ee
These relations  reveal the PHD. Next we will discuss the physical consequences of the duality.

%%%%%%%%%%%%%%%%%%
\paragraph{Particle-hole duality.}
%%%%%%%%%%%%%%%%%%

The Hall conductivity and curvature response in the PH-dual state can be found using \eqref{density}. 
Taking a derivative w.r.t. the magnetic field $B(q)$, and applying the St\v reda formula we obtain an exact relation between the Hall conductivities
\be\la{sigma1}
\sigma_\nu^H(q) + \sigma_{1-\nu}^{H}(q) = \sigma_1^{H}(q)\,.
\ee
Similarly we find
\be\la{eta1}
\nu \,\eta_{\nu}(q) + (1-\nu)\eta_{1-\nu}(q) = \eta_1(q)\,.
\ee
Next, we turn to the normalization factor. It follows directly from the definition of the reduced density matrix and the reproducing formula \eqref{projector}, as well as the definition of the generating functional \eqref{genfun_nu} that
\be
\label{exactZ}
\frac{\cZ_{1-\nu}}{\mathcal{Z}_{\nu}} = \mathcal{Z}_{1}\quad \Rightarrow \quad  \log\cZ_{1-\nu} - \log \cZ_\nu = \log \cZ_1\,\,,
\ee
where we have dropped the argument of $\cZ_\nu$ for brevity.
Eq.\eqref{exactZ} is an {\it exact} relation between the generating functionals for a pair of PH-dual states. Eq.\eqref{exactZ} clearly illustrates the duality. 

Variation of $\log \cZ_{1-\nu}$ over $W(\zeta)$ is given by
\be\la{firstD}
\frac{\delta}{\delta W(\zeta)} \log \cZ_{1-\nu} = \langle \rho_1\rangle  + \langle \rho_\nu \rangle = \langle \rho_{1-\nu} \rangle  + 2\langle \rho_\nu \rangle  \,.
\ee
We emphasize that since \eqref{firstD} does not have the same form as \eqref{GenF}, the wavefunction $\Psi_{1-\nu}$ does not have the form \eqref{ansatzWF}. More precisely, we see that $f_{1-\nu}$ {\it has to} depend on $W$. In other words, the dual states couple differently to inhomogeneous magnetic field. It appears that the condition for $f_\nu$ to be independent of $W$ has to do with the chirality of a state -- for instance, all conformal block trial states share this property.  This complication indicates that identity \eqref{exactZ} is not sufficient to extract all of the observables in a PH-dual state in terms of the observables in the original state, because the relationship between the observables and variations of $\log\cZ_{1-\nu}$ is more complicated for $\Psi_{1-\nu}$ states.

Now we will derive an analogue of \eqref{SSFsigma} for the dual states. We utilize the fact that, by definition, the two-point density correlation function is related to the 2-particle density matrix via
$$ 
\langle \rho_\nu(\zeta) \rho_\nu(\zeta^\prime) \rangle = \langle \rho_\nu(\zeta)\rangle \delta(\zeta - \zeta^\prime) + 2\cP_\nu^{(2)}(\zeta,\zeta^\prime; \zeta,\zeta^\prime)\,.
$$
This allows us to gain insight using the exact formula \eqref{exact_density}. Then the one-particle density matrix for the $\nu$ state is known in the translation-invariant limit to be \cite{macdonald1988density}
\begin{align}
\cP_{\nu}^{(1)}(\zeta; \zeta') = \bar{\rho}_{\nu}	e^{\zeta \bar{\zeta}'/2l^{2} - |\zeta'|^{2}/2\ell^{2}}.
\end{align}
Using this in \eqref{exact_density}, we find 
\be\label{trans_inv}
\langle \rho_{1-\nu}(\zeta)\rho_{1-\nu}(\zeta^\prime)\rangle_{c} = \langle \rho_\nu(\zeta)\rho_\nu(\zeta^\prime)\rangle_{c}+\frac{\bar\rho_1-2\bar \rho_\nu}{\bar \rho_1}\langle \rho_1(\zeta)\rho_1(\zeta^\prime)\rangle_{c}\,.
\ee
Taking the Fourier transform, we find a beautiful exact relation between the projected static structure factors $\bar{s}_{\nu} = s_{\nu} - s_{1}$ for a pair of PH-dual states \footnote{Levin and Son have obtained the same relation using a different set of arguments \cite{SL}}
\be\la{sonstrick}
\bar \rho_\nu \bar s_\nu(q) = \bar \rho_{1-\nu} \bar s_{1-\nu}(q)\,.
\ee

Now we are in position to relate the Hall conductivity to the SSF of the PH-dual state
%\be\la{SSFsigmaPH}
% \sigma^{1-\nu}_H(q)= \frac{2(1-\nu)}{2\pi q^2}\Big(s_1(q) - \bar s_{1-\nu}(q)\Big)\,.
%\ee
\be\la{SSFsigmaPH}
\sigma_{1-\nu}^H(q)= \sigma_{1-\nu}^{H}(0) \frac{2}{q^2}\Big( s_{1}(q) -\bar s_{1-\nu}(q)\Big)\,.
\ee
 The simplest way to obtain \eqref{SSFsigmaPH} is to use \eqref{sigma1}, \eqref{SSFsigma} and \eqref{sonstrick}.
 
 Next we will derive an analogue of \eqref{SSFeta} for PH-dual states. Using \eqref{sigma1}, \eqref{eta1} and \eqref{sigma-eta} we find 
\be\la{SSFetaPH}
\bar s_{1-\nu} (q) = s_1(q) - \frac{q^2/2}{1+q^2/2}(q^2  \eta_{1-\nu}(q) +1)\,.
\ee
Excluding $s_{1-\nu}(q)$ from \eqref{SSFsigmaPH}-\eqref{SSFetaPH}  we come to a surprising conclusion -- the relation between $\sigma^H_{1-\nu}(q)$ and $\eta_{1-\nu}(q)$ is precisely the same as before the PH-transfomation \eqref{sigma-eta}, up to replacing $\nu$ by $1-\nu$.

\paragraph{Berry curvature.}

Next we turn to the dependence of the PH-dual states on parameters such as adiabatically varying fluxes of magnetic field or the modular parameter of a torus $\tau$. Denote any of these parameters in complex coordinates as $x$ and $\bar x$. Berry curvature can be computed under the assumption that the state $\Psi_{\nu}$ is holomorphic in the coordinates on the parameter space, except for the real-analytic normalization factor.  The {\it normalized} states have the form \cite{2009-Read-HallViscosity, bradlyn2015topological,klevtsov2015precise}
\begin{align}
\psi_\nu\left(\{\xi, \bar{\xi}\}; x, \bar x\right) = \frac{1}{\sqrt{\cZ_\nu[x, \bar{x}]}} \Psi_\nu\left(\{\xi, \bar{\xi}\}; x\right)\,.
\end{align}
Then the holomorphic component of the Berry connection is determined entirely by the normalization factor
\begin{align}
A_{x} \equiv i \langle \psi_{\nu} | \partial_{x} | \psi_{\nu}\rangle = \frac{i}{2} \partial_{x} \log \mathcal{Z}_{\nu},
\end{align}
which follows by using the identity  $\partial_{x} \langle \psi_{\nu} | \psi_{\nu} \rangle = 0$ to trade derivatives of $\Psi_\nu$ for derivatives of $\mathcal{Z}_\nu$. Thus, for such holomorphic states the Berry curvature is a  K\"ahler form with the K\"ahler potential $\mathcal{U}_{\nu} = \log \mathcal{Z}_{\nu}$, and is given by
\begin{align}
\Omega_{\nu} = \frac{i}{2}\left( \partial_{x} \partial_{\bar{x}} \mathcal{U}_{\nu}\right)	\,d x \wedge d\bar{x}\,.
\end{align}
This structure is nearly preserved for the PH-dual state. A straightforward calculation shows that the K\"ahler potential is $\mathcal{U}_{1-\nu} = \log (\cZ_1/\cZ_\nu)$, which is {\it not} the logarithm of the normalization as before. Thus, in contrast to the formula \eqref{exactZ}, the Berry curvature obeys 
\be\label{Berry}
 \Omega_{\nu} +\Omega_{1-\nu}  = \Omega_{1}.
\ee

%%%%%%%%%%%%%%%%%%%%%%%%%%%%
\paragraph{PHD and Chern-Simons terms.} 
%%%%%%%%%%%%%%%%%%%%%%%%%%%%

The first few terms in the long wave expansion of $\sigma_\nu^H(q)$, $s_\nu(q)$ and $\eta_\nu(q)$ are determined by the topological quantum numbers, which appear as the coefficients of the Chern-Simons terms in the effective action \cite{bradlyn-read-2012kubo, CLW, Gromov-galilean}.

The Chern-Simons part of the effective action is given by \cite{GCYFA, bradlyn2015topological}
\be
\mathcal W^\nu_{\rm CS} = \frac{\nu}{4\pi} \int \Big( A + \bar s \omega\Big) d \Big(A + \bar s \omega\Big)  - \frac{\tilde c}{48\pi}\int \omega d \omega\,,
\ee
where $\tilde c=c_- -  12\nu \,{\rm var}(s)  $ and other coefficients are the topological quantum numbers discussed in the introduction. We have also introduced $\omega_\mu$ -- a spatial part of the spin connection satisfying $\p_1 \omega_2 - \p_2\omega_1 = \frac{1}{2} \sqrt{g} R$. This effective action encodes the linear response functions. Notably, the Hall conductance, shift, and the Hall viscosity, averaged over the sample, are given by
\be
\sigma^H_\nu = \frac{\nu}{2\pi}~,\quad \mathcal S =\frac{\nu^{-1} N - N_\phi}{\chi/2}~,\quad  \eta^H_\nu = \frac{\bar s}{2} \bar \rho_\nu + \frac{\tilde c}{24} \frac{\chi}{A}\,,
\ee
where $A$ and $\chi$ are the area and the Euler characteristic of the sample, $N$ is the number of electrons and $N_\phi$ is the total magnetic flux in units of the flux quantum. When a FQH state is constructed as a single conformal block in a conformal field theory \cite{moore1991nonabelions, bradlyn2015topological}  $\tilde c = c_-$. However, in general (and, notably, for Jain states) ${\rm var}(s)$ does not vanish \cite{GCYFA}.

The action of PH-transformation on the Chern-Simons part of the effective action is
\be
\mathcal W^\nu_{\rm CS}  + \mathcal W^{1-\nu}_{\rm CS} =\mathcal W^{1}_{\rm CS}\,.
\ee
This can be seen as a consequence of the formula for the Berry curvature \eqref{Berry} following the arguments of \cite{bradlyn2015topological}. In addition to $\nu^{PH}= 1-\nu$ it implies 
\be\la{PHmap}
\quad \cS^{PH}  = \frac{1-\nu \cS}{1-\nu}\,,\quad \vars^{PH} = \frac{\nu}{\nu-1} \left(\frac{ (1-\cS)^2}{4(1-\nu)} + \vars\right)\,.
\ee
PHD also transforms the chiral central charge according to $c_-^{PH} = 1 - c_-$, and  $\tilde c^{PH} = c_{-}^{PH} - 12 \nu^{PH} \vars^{PH}$. Curiously, if the initial state had $\vars=0$, then $\vars^{PH} \neq 0$, unless $\cS=1$.

These relations are in agreement with the results derived above. For example, using the transformation laws \eqref{PHmap}, combined with \eqref{buniversal} we can check that relations \eqref{sigma1} and \eqref{eta1} hold at the order $q^4$ and $q^2$ correspondingly.

As another example we provide an explicit formula for first $2$ terms in the long-wave expansion of the projected SSF of a PH-dual state
\bea
\!\!\!\!\!\bar s_{1-\nu}(q) = \frac{\nu (\cS-1)}{8 (1- \nu)} q^4 + \frac{  (-6 b +5 \nu -3 \nu  \cS)}{48 (1-\nu)}q^6 + \ldots\,,
\eea
where $b$ is given by \eqref{buniversal} and all of the topological quantum numbers are known for a large variety of states \cite{GCYFA}. Note that all of the quantum numbers are taken from the state at filling $\nu$. We are not aware of this type of general result in the literature.

%%%%%%%%%%%%%%%%%%%%%%%%%%%%
\paragraph{Jain states.} 
%%%%%%%%%%%%%%%%%%%%%%%%%%%%

We apply our relations to the PH-duals of $\nu = \frac{N}{2N+1}$ Jain states. Other quantum numbers are given by
\be
\mathcal S = N+2\,,\quad c_{-}=N\,, \quad \nu\,\vars = \frac{N(N^2-1)}{12}\,.
\ee
Then the projected SSF of the PH-dual state is given by
\be\la{SSFj}
\bar s_{\frac{N+1}{2N+1}}(q) =\frac{N}{8} q^4+\frac{N^4+2 N^3-2 N^2-2N}{48 (N+1)}q^6 + \ldots\,.
\ee
To the best of our knowledge \eqref{SSFj} is a new result.

Hall conductivity of Jain's state at the filling factor $\nu=\frac{N}{2N+1}$ and its PH-dual state at  the filling factor $(1-\nu)=\frac{N+1}{2N+1}$ can be calculated in closed form in the large $N$ limit, so that $z=q(2N+1)\sim 1$ \cite{NS}. In the absence of long-range interactions
\begin{equation}\la{sigmaJain}
\sigma^H_\nu(q)=\frac{((4 N+2)^2-z^2) \left(8 N+\frac{2 z J_2(z)}{J_1(z)}\right)}{64 \pi  (2 N+1)^3}\,,
\end{equation}
\begin{equation}\la{sigmaJainM}
\sigma^H_{1-\nu}(q)=\frac{((4 N+2)^2-z^2) \left(8 N+8-\frac{2 z J_2(z)}{J_1(z)}\right)}{64 \pi  (2 N+1)^3}\,.
\end{equation}
where $J_\alpha(z)$ is the Bessel function. The correction to this is order $O(N^{-4})$. 

The projected SSF can also be derived in closed form using Dirac composite fermion theory \cite{son2015composite}
\bea\la{sJain}
&&\bar{s}_\nu(q)=\frac{z^3 ((4 N+2)^2-z^2) J_2(z)}{32 N (2 N+1)^4 J_1(z)}\,,
\\ 
\la{sJainM}
&&\bar{s}_{1-\nu}(q)=\frac{z^3 ((4 N+2)^2-z^2)  J_2(z)}{32 (N+1) (2 N+1)^4 J_1(z)}\,.
\eea
With these expressions at hand we can check that \eqref{sigma1} holds up to order $N^{-2}$. % for all $q$ in the leading and sub-leading order in $N$.
It also follows from \eqref{sigmaJain}-\eqref{sJainM} that \eqref{SSFsigma} and \eqref{SSFsigmaPH} hold in the large $N$ limit at leading and sub-leading orders in $N$. We emphasize that these are quite non-trivial checks that probe the relations we derived in {\it all} orders in the momentum expansion.

Long-range interactions can be included in the computation of \cite{NS}, which leads to the breakdown of relations \eqref{SSFeta} and \eqref{SSFetaPH} at $O(q^7)$ order and of \eqref{sigma-eta} at $O(q^3)$ order.

%%%%%%%%%%%%%%%%%%%%%%%%%%%%
\paragraph{Conclusion.} 
%%%%%%%%%%%%%%%%%%%%%%%%%%%%
We have presented arguments for the particle-hole duality in the lowest Landau level. This duality implies several exact, non-perturbative relations between the observables in the pair of PH-dual states such as static structure factor, Hall conductivity and response of the electron density to curvature. Our results do not cover all possible states in the lowest Landau level -- it is possible to start with a non-chiral state so that its PH-dual is also non-chiral. PH-Pffafian is an example of such a state, however \eqref{sigma-eta} and \eqref{sonstrick} should be applicable to such states as well. We leave the investigation of general non-chiral states for the future work.

%%%%%%%%%%%%%%%%%%%%%%%%%%%%
\paragraph{Acknowledgements} 
%%%%%%%%%%%%%%%%%%%%%%%%%%%%

 It is a pleasure to thank D.T. Son and P. Wiegmann for very enlightening discussions and correspondence.

D.X.N. is supported by the Chicago MRSEC, which is
funded by NSF through grant DMR-1420709.
A.G. was supported by the Leo Kadanoff Fellowship.
\bibliography{Bibliography}
\newpage
\appendix
\begin{widetext} 

\section{Supplementary Material}

\subsection{Lowest Landau level states}

The orthonormal complete basis set of single particle wave functions in the  LLL is
\begin{equation}
\phi_m(z)=\frac{z^m}{\sqrt{2\pi 2^m m!}}e^{-\frac{1}{4\ell^2}|z|^2}\,.
\end{equation}
We will take $\ell=1$ to simplify the formulas. We have the projected annihilation field operator
\begin{equation}
\psi_L(z)=\sum_{m=0}^{\infty}\hat{a}_m \phi_m(z)=\sum_{m=0}^{\infty}\hat{a}_m \frac{z^m}{\sqrt{2\pi 2^m m!}}e^{-\frac{1}{4}|z|^2}\,,
\end{equation}
where $\hat{a}_m$ and $\hat{a}^\dagger_m $  annihilate and create normalized LLL states and obey the Fermi canonical relation
\begin{equation}
\label{eq:antcom}
\lbrace \hat{a}_n,\hat{a}^\dagger_m\rbrace=\delta_{mn}\,, \qquad \lbrace \hat{a}^\dagger_n,\hat{a}^\dagger_m\rbrace=\lbrace \hat{a}_n,\hat{a}_m\rbrace=0\,.
\end{equation}
Thus equal time anti-commutators of projected electron field operators are
\begin{equation}
\lbrace \psi_L^\dagger(z_1),\psi_L(z_2)\rbrace = \sum_{m=0}^{\infty}\phi^*_m(z_1)\phi_m(z_2)=\frac{1}{2\pi}e^{-\frac{1}{4}|z_1-z_2|^2}e^{\frac{1}{4}(\bar{z}_1z_2-\bar{z}_2 z_1)}\equiv\lbrace z_1 | z_2 \rbrace. 
\end{equation}
Since the LLL single-particle states are not complete in the {\it full} Hilbert space $\{ z_1|z_2\} \neq \delta^2(z_1-z_2)$. In fact, $\lbrace z_1 | z_2 \rbrace$ acts as the $\delta$-function in the LLL  
\begin{equation}
\int d^2 z_1 F(z_1) \lbrace z_1 | z_2 \rbrace = F(z_2)\,,
\end{equation}
where $F(z)$ is any function of the form $F(z)=f(z)e^{-\frac{1}{4}|z|^2}$ and $f(z)$ is holomorphic. 
Further more,
\begin{equation}
\int d^2 z_2  \lbrace z_1 | z_2 \rbrace G(\bar{z}_2) = G(\bar{z}_1)\,,
\end{equation}
where $G(\bar{z})$ is any function of the form $G(\bar{z})=g(\bar{z})e^{-\frac{1}{4}|z|^2}$ and $g(\bar{z})$ is anti-holomorphic. 

Finally, $\{ z_1|z_2\}$ satisfies the composition rule
\be
\int d^2 z_2 \lbrace z_1 | z_2 \rbrace \lbrace z_2 | z_3 \rbrace=\lbrace z_1 | z_3 \rbrace\,,\qquad \lbrace z_1 | z_2 \rbrace^*=\lbrace z_2 | z_1 \rbrace\,.
\ee

\subsection{LLL at finite number of particles}

Next we assume that only $N$ orbitals are available in the LLL. The finite $N$ version of the LLL $\delta$-function is
\begin{equation}
K_N(z_1; z_2)=\sum_{m=0}^{N-1}\phi_m(z_1)\phi^*_m(z_2)\,,
\end{equation}
which is known as the reproducing kernel for the Hilbert space of states spanning the LLL (also known as the Bergman kernel), which satisfies the composition rule
\begin{align}
K_{N}(z; z') = \int d^{2} \xi K_{N}(z; \xi) K_{N}(\xi; z'),	
\end{align}
%Furthermore, it acts as a $\delta$-function for  polynomials constructed from 
and the reproducing formula %properties that 
\begin{equation}
\label{eq:F}
\int d^2 z_2  K_{N}( z_1 ; z_2) F_N(z_2)  = F_N(z_1)\,,
\end{equation}
where $F_{N}(z)$ is any linear combination of $\phi_m(z)$ with $m$ running from $0$ to $N-1$. 
Similarly,
\begin{equation}
\label{eq:G}
\int d^2 z_1  G_N(\bar{z}_1)K_{N}(z_{1}; z_{2})   = G_N(\bar{z}_2)\,,
\end{equation}
where $G_N(\bar{z})$  is any linear combination of $\phi^*_m(z)$ with $m$ running from $0$ to $N-1$.

The normalized wave function of the $\nu=1$ integer quantum Hall state is 
\begin{eqnarray}
\Psi_1(z_1,\cdots,z_N)=\frac{1}{\sqrt{N!}}\begin{vmatrix}
\phi_0(z_1) &\phi_0(z_2)  & \cdots & \phi_0(z_N) \\ 
\phi_1(z_1) &\phi_1(z_2)  & \cdots & \phi_1(z_N) \\
\cdots & \cdots & \cdots & \cdots \\ 
\phi_{N-1}(z_1) &\phi_{N-1}(z_2)  & \cdots & \phi_{N-1}(z_N) \\ 
\end{vmatrix}\,.
\end{eqnarray}
We define the $n-$particle density matrix for this state by

\be
\label{eq:RM0}
\cP_{1}^{(n)}(\cxi,\cxip) = \frac{N!}{(N-n)! n!} \int \dz \Psi_{1}(\cz,\cxi)\Psi^*_{1}(\cz,\cxip)\,.
\ee

%\be
%\label{eq:RM0}
%\cP_{N}^{(k)}(\cxi,\cxip) = \frac{N!}{(N-k)! k!} \int \dz \Psi_1(\cz,\cxi)\Psi^*_1(\cz,\cxip)\,.
%\ee
where we used the shorthand notation
\begin{equation}
	\cz=z_1,\cdots,z_M \qquad \cxi=z_{M+1},\cdots,z_{N} \qquad \cxip=z^\prime_{M+1},\cdots,z^\prime_{N}.
\end{equation} 
with $n = N - M$. It can be  shown that
\begin{align}
\cP_{1}^{(n)}(\xi_{1}, ..., \xi_{n}; \xi_{1}', ..., \xi_{n}') = \frac{1}{n!}\det \left[ K_{N}(\xi_{i}; \xi_{j}') \right]_{1 \le i,j \le n}\,\,.
\end{align}

For example, $\cP_{1}^{(1)}(\xi; \xi')= K_{N}(\xi; \xi')$, which at equal points coincides with the mean density of the integer state  $\langle \rho_{1}(\xi)\rangle  = K_{N}(\xi; \xi)$. Furthermore, it's clear that
\begin{align}
\cP_{1}^{(2)}(\xi_1, \xi_2; \xi_1', \xi_2') = \frac{1}{2}(K_{N}(\xi_1;\xi_1') K_{N}(\xi_2;\xi_2') - K_{N}(\xi_1; \xi_2') K_{N}(\xi_2;\xi_1'))\,.
\end{align}

On the diagonal, $\cP^{(2)}(\xi_1, \xi_2; \xi_1, \xi_2)$ is related to the two-particle distribution function $\cP^{(2)}(\xi_1 , \xi_2; \xi_1, \xi_2) = \frac{1}{2} n^{(2)}(\xi_{1}, \xi_{2})$, which can be expressed in terms of the pair distribution function $n^{(2)}(\xi_{1}, \xi_{2}) = \langle \rho (\xi_1)\rangle \langle\rho (\xi_2)\rangle g(\xi_1, \xi_2)$. Similar to the reproducing kernel, the reduced density matrix has the property that
\begin{align}
\label{eq:norm}
\int [d\xi'] \cP_{1}^{(n)}(\{\xi\}; \{\xi'\}) F_{N}(\{\xi'\}) =  F_{N}(\{\xi\})\,.	
\end{align}
for any {\it antisymmetric} function $F_{N}(\{\xi\}) = F_{N}(\xi_{1}, ..., \xi_{n})$ built by linear superposition of the $N$ LLL orbitals. In particular, for a fermionic  $M-$ particle wave function $\Psi_{\nu}$, it clearly follows that
\begin{align}
\int [d\xi'] \cP_{1}^{(M)}(\{\xi\}; \{\xi'\}) \Psi_{\nu}(\{\xi'\}) = \Psi_{\nu}(\{\xi\})\,.	
\end{align}
Then defining PH conjugation by
\begin{align}
\Psi_{1-\nu}(\{z\})  = \sqrt{ \frac{N!}{(N-M)! M!}} \int [d\xi] \Psi_{1}(\{z\}, \{\xi\}) \Psi_{\nu}^{*}(\{\xi\})\,.	
\end{align}
we find
\begin{align}
\Psi_{1-\nu}^{PH} = \Psi_{1-(1-\nu)} = (-1)^{M(N-M)} \Psi_{\nu}	.
\end{align}

Results of this Section remain true in inhomogeneous magnetic field and curvature.

%%%%%%%%%%%%%%%%%%%%%%%%%%%%%%%%%%%%%%%%%%%%%%%%%%%%%%%%
%%%%%%%%%%%%%%%%%%%%%%%%%%%%%%%%%%%%%%%%%%%%%%%%%%%%%%%%
%
%	\begin{align}
%	\Psi_{1-\nu}^{PH} &= \sqrt{\frac{N!}{(N - M)! M!	}} \int [d\xi]_{N-M} \Psi_{1}([z]_M, [\xi]_{N-M}) \Psi_{1-\nu}^{*}([\xi]_{N-M})\\
%	& = \frac{N!}{(N-M)! M!}\int [d\xi]_{N-M} [d\xi']_{M} \Psi_{1}([z]_{M}, [\xi]_{N-M}) \Psi_{1}^{*}([\xi]_{N-M}, [\xi']_{M}) \Psi_{\nu}([\xi']_{M})
%	\end{align}

%To get from 

%\begin{align}
%\Psi_{1}([z]_{M}, [\xi]_{N-M} ) \to \Psi_{1}( [\xi]_{N-M}, [z]_{M}) (-1)^{M(N-M)	}
%\end{align}

%So 
%\begin{align}
%\Psi_{1-\nu}^{PH} = (-1)^{M(N-M)} \int [d\xi']_{M} \cP_{N}^{(M)}([z]; [\xi']	) \Psi_{\nu}([\xi']_{M}) = (-1)^{M(N-M)} \Psi_{\nu}
%\end{align}

%%%%%%%%%%%%%%%%%%%%%%%%%%%%%%%%%%%%%%%%%%%%%%%%%%%%%%%%
%%%%%%%%%%%%%%%%%%%%%%%%%%%%%%%%%%%%%%%%%%%%%%%%%%%%%%%%

\subsection{PHD for the density matrix}

Here we obtain a relation between the density matrices computed for different filling fractions. Specifically
\begin{align}
\cP_{\nu}^{(n)}(\{\xi\}; \{\xi'\}) = \frac{N_{\nu}!}{(N_{\nu}- n)! n!} \int [dz] \Psi_{\nu}(\{z\}, \{\xi\}) 	\Psi_{\nu}^*(\{z\}, \{\xi'\}) 
\end{align}
where $N_{\nu} = M$ is the total number of particles in the state $\Psi_{\nu}$. It is similarly defined for the PH-conjugate state $\Psi_{1-\nu}$, with $N_{1-\nu}= N-M$. We start with the one-particle density matrix 
\begin{align}
\cP_{1-\nu}^{(1)}(z; z'	)& \equiv  \frac{ (M+1)!}{M!} \int [d\xi] [d\xi'] \cP_{1}^{(M+1)}(z, \{\xi\}; z', \{\xi'\}) \Psi_{\nu}(\{\xi'\}) \Psi_{\nu}^{*}(\{\xi\}), \label{1ptden_def}\\
& = \cP_{1}^{(1)}(z; z') - \cP_{\nu}^{(1)}(z; z') \label{1ptden}\,,
\end{align}
where the second equality follows directly from the following expansion of the determinant in the definition of the density matrix
\begin{align}
\cP_{1}^{(M+1)}(z, \{\xi\}; z', \{\xi'\}) = \frac{1}{M+1} \left[ K_{N}(z; z') \cP_{1}^{(M)}(\{\xi\}, \{\xi'\}) - \sum_{k = 1}^{M} K_{N}(z; \xi_{k}') \cP_{1}^{(M)}(\xi_{k}, \{\xi\}_{k}; z', \{\xi'\}_{k})	 \right]\,,
\end{align}

where $\{\xi\}_{k} = \{\xi_{1}, ..., \xi_{k-1}, \xi_{k+1}, ..., \xi_{M}\}$ is the ordered array of coordinates with $\xi_{k}$ excluded. On the diagonal, $\cP^{(1)} = \langle \rho \rangle$, and  \eqref{1ptden} implies the duality relation for the mean density $\langle \rho_{1-\nu}\rangle = \langle \rho_{1}\rangle - \langle \rho_{\nu}\rangle$. 

Next, we consider the two-particle reduced density matrix, which can be written as a convolution involving the $(M+2)$-particle density matrix
\begin{align}
\cP_{1-\nu}^{(2)}(z_1, z_2; z_1, z_2)
& = \frac{(M+2)!}{2 (M!)} \int [d\xi] [d\xi'] \cP_{1}^{(M+2)}(z_1, z_2, \{\xi\}; z_1, z_2, \{\xi'\}) \Psi_{\nu}(\{\xi'\}) \Psi_{\nu}^{*}(\{\xi\})
\end{align}
We next utilize the expansion
\begin{align}
\cP_{1}^{(M+2)}(z_1, z_2, \{\xi\}; z_1, z_2, \{\xi'\}) = \frac{1}{M+2} \Big[ &K_{N}(z_1; z_1) \cP_{1}^{(M+1)}(z_2, \{\xi\}; z_2 , \{\xi'\}) \nonumber\\
&-  K_{N}(z_1; z_2) \cP_{1}^{(M+1)}	(z_2, \{\xi\}; z_1, \{\xi'\})\nonumber\\
& - \sum_{k = 1}^{M} (-1)^{k}K_{N}(z_1; \xi_{k}') \cP_{1}^{(M+1)}( z_2, \{\xi\}; z_1, z_2, \{\xi'\}_{k}) \Big]\label{exp3}
\end{align}
Inserting this into the definition and using \eqref{1ptden_def}, we get
\begin{align}
\cP_{1-\nu}^{(2)}(z_1, z_2; z_1, z_2) &  = \frac{1}{2} \Big[K_{N}(z_1; z_1)	\cP_{1-\nu}^{(1)}(z_2; z_2) - K_{N}(z_1; z_2) \cP_{1-\nu}^{(1)}(z_2; z_1)\nonumber \\
 &  - \frac{1}{2} (M+1) \sum_{k} (-1)^{k} \int [d \xi'] [d\xi] K_{N}(z_1; \xi_{k}') \cP_{1}^{(M+1)}(z_{2}, \{\xi\}; z_1, z_2, \{\xi'\}_{k}) \Psi_{\nu}(\xi_{k}', \{\xi'\})\Psi_{\nu}^{*}(\xi_{k}, \{\xi\}_{k})\,.
\end{align}

%Then we may use

%\begin{align}
%\int [d\xi] [d\xi'] \cP_{1}^{(M+1)} (z, \{\xi\}; z', \{\xi'\}) \Psi_{\nu}(\xi') \Psi_{\nu}^{*}(\xi) = \frac{1}{M+1} \cP_{1-\nu}^{(1)}(z; z') 	
%\end{align}

In order to evaluate the sum in the second line, we use the expansion

%\begin{align}
%\cP_{1}^{(M+1)}(z_2, \{\xi\}; z_1, z_2, \{\xi'\}_{k}) = \frac{1}{(M+1)} \Big[ &K_{N}(z_2; z_1) \cP_{1}^{(M)}(\{\xi\}; z_2, \{\xi'\}_{k})\\
%& - K_{N}(z_{2}, z_{2}) \cP_{1}^{(M)}(\{\xi\}; z_{1}, \{\xi'\}_{k})\\
%& + K_{N}(z_{2}, \xi_{1}') \cP_{1}^{M}(\{\xi\}; z_{1}, z_{2}, \{\xi'\}_{1,k} )\\
%& ... \Big]
%\end{align}

\begin{align}
\cP_{1}^{(M+1)}(z_2, \{\xi\}; z_1, z_2, \{\xi'\}_{k}) = \frac{1}{(M+1)} \Big[ &K_{N}(z_2; z_1) \cP_{1}^{(M)}(\{\xi\}; z_2, \{\xi'\}_{k})\nonumber\\
& - K_{N}(z_{2}, z_{2}) \cP_{1}^{(M)}(\{\xi\}; z_{1}, \{\xi'\}_{k})\nonumber\\
& + \sum_{j \ne k = 1}^{M} (-1)^{j+1}K_{N}(z_{2}, \xi_{j}') \cP_{1}^{(M)}(\{\xi\}; z_{1}, z_{2}, \{\xi'\}_{j,k} )\Big]\,.
\end{align}

Plugging this in and using \eqref{eq:norm} gives

\begin{align}
\cP_{1-\nu}^{(2)}(z_1, z_2; z_1, z_2) &  = \frac{1}{2} \Big[K_{N}(z_1; z_1)	\cP_{1-\nu}^{(1)}(z_2; z_2) - K_{N}(z_1; z_2) \cP_{1-\nu}^{(1)}(z_2; z_1)\nonumber \\
 &  - \frac{1}{2}  \sum_{k} (-1)^{k} \int [d \xi'] K_{N}(z_1; \xi_{k}') K_{N}(z_2; z_1)  \Psi_{\nu}(\{\xi'\})\Psi_{\nu}^{*}(z_2, \{\xi'\}_{k})\nonumber\\
  &  + \frac{1}{2}  \sum_{k} (-1)^{k} \int [d \xi'] K_{N}(z_1; \xi_{k}')  K_{N}(z_2; z_2) \Psi_{\nu}(\{\xi'\})\Psi_{\nu}^{*}(z_1, \{\xi'\}_{k})\nonumber\\
  & - \frac{1}{2} \sum_{k \ne j} (-1)^{k + j + 1} \int [d\xi'] K_{N}(z_1; \xi_{k}') K_{N}(z_2; \xi_{j}') \Psi_{\nu}(\{\xi'\}) \Psi_{\nu}^{*}(z_1, z_2, \{\xi'\}_{j,k})\Big] \,.
\end{align}

which can be expressed in terms of the reduced density matrix of the $\Psi_{\nu}$ state as 
\begin{align}
\cP_{1-\nu}^{(2)}(z_1, z_2; z_1, z_2) &  = \frac{1}{2} \Big[K_{N}(z_1; z_1)	\cP_{1-\nu}^{(1)}(z_2; z_2) - K_{N}(z_1; z_2) \cP_{1-\nu}^{(1)}(z_2; z_1) \nonumber\\
 &  + \frac{1}{2}  \int d^{2}\xi_{1}' K_{N}(z_1; \xi_{1}') K_{N}(z_2; z_1)  \cP_{\nu}^{(1)}(\xi_{1}'; z_{2})\nonumber\\
  &  - \frac{1}{2}   \int d^{2}\xi_{1}' K_{N}(z_1; \xi_{1}')  K_{N}(z_2; z_2) \cP_{\nu}^{(1)}(\xi_{1}'; z_{1})\nonumber\\
  & +  \int d^{2} \xi_{1}' d^{2} \xi_{2}' K_{N}(z_1; \xi_{1}') K_{N}(z_{2}; \xi_{2}') \cP_{\nu}^{(2)}(\xi_{1}', \xi_{2}'; z_{1}, z_{2})\Big] 
\end{align}

using the reproducing property of the kernel, and the determinant expression for the 2-particle density matrix $\cP_{1}^{(2)}$, this can be brought into the form presented in the paper. 

\subsection{Two-point functions of density from the generating functional}

Next we explain how to obtain connected two-point functions of density $\rho_{1-\nu}$ from the generating functional  $\log \cZ_{1-\nu}$. Indeed, differentiating  \eqref{exactZ} w.r.t. $W(\zeta^\prime)$
\be\la{1minnu}
\frac{\delta}{\delta W(\zeta^\prime)} \langle \rho_{1-\nu}(\zeta)\rangle = \langle \rho_{1-\nu}(\zeta^\prime) \rho_{1-\nu}(\zeta)\rangle_{c} + [\rho_{\nu}(\zeta^\prime);\rho_{1-\nu}(\zeta)]\,,
\ee

where we have introduced a ``mixed'' correlator
 
\be\la{mixed}
 [\rho_{\nu}(\zeta^\prime);\rho_{1-\nu}(\zeta)]=2\frac{(N+M)!}{N!M!}\int \dxi \dxip \Psi_\nu(\cxi) \Psi_\nu^*(\cxip){\rho}_\nu(\zeta^\prime) \int \dz \Psi_1(\cz,\cxip) \Psi_1^*(\cz,\cxi){\rho}_{1-\nu}(\zeta)\,,
\ee
which signals the failure of the formula \eqref{2ptden} for PH-dual states. To evaluate this term we combine \eqref{1minnu} together with \eqref{density} and \eqref{firstD} to obtain
\bea\nonumber
\frac{\delta^2 \ln \cZ_{1-\nu} }{\delta W(\zeta^\prime)\delta W(\zeta)}  &&=\langle \rho_{1-\nu}(\zeta)\rho_{1-\nu}(\zeta^\prime) \rangle_{c} + 2\langle \rho_\nu(\zeta) \rho_\nu(\zeta^\prime) \rangle_{c} +  [\rho_{\nu}(\zeta^\prime);\rho_{1-\nu}(\zeta)]\\
&&=  2 \langle \rho_1(\zeta) \rho_1(\zeta^\prime) \rangle_{c} - \langle \rho_{1-\nu}(\zeta)\rho_{1-\nu}(\zeta^\prime) \rangle_{c}  -  [\rho_{\nu}(\zeta^\prime);\rho_{1-\nu}(\zeta)]\,.
\eea
Which implies \footnote{If the connected two-point functions were related in a manner similar to the one-point functions \eqref{density}, then the mixed correlator would vanish. }
\begin{align}\la{trick}
 [\rho_{\nu}(\zeta^\prime);\rho_{1-\nu}(\zeta)] = &\langle \rho_1(\zeta) \rho_1(\zeta^\prime)\rangle_{c}	-\langle \rho_{1-\nu}(\zeta)\rho_{1-\nu}(\zeta^\prime\rangle_{c}- \langle \rho_\nu(\zeta)\rho_\nu(\zeta^\prime) \rangle_{c} \,.
\end{align}

In the translation-invariant state, we can use \eqref{trans_inv} to evaluate this in momentum space, and combined with \eqref{sonstrick} 
we find a simple expression for the mixed correlator
\be\la{mixedfinal}
[\rho_{\nu};\rho_{1-\nu}](q)=- 2\bar \rho_\nu \bar s_\nu =-2\bar \rho_{1-\nu} \bar s_{1-\nu}\,.
\ee

\end{widetext}

\end{document}